\documentclass[aoas,MSNbibl,nameyear,dvips]{arximspdf}
\usepackage{graphicx}
\doi{10.1214/12-AOAS615} 
\volume{7}
\issue{2}
\pubyear{2013}
\firstpage{989}
\lastpage{1009}


\begin{document}
\begin{frontmatter}

\title{Bayesian alignment of similarity shapes}
\runtitle{Bayesian similarity shape}

\begin{aug}
\author[E]{\fnms{Kanti V.} \snm{Mardia}\ead[label=e1]{mardia@stats.ox.ac.uk}},
\author[B]{\fnms{Christopher J.} \snm{Fallaize}\ead[label=e2]{Chris.Fallaize@nottingham.ac.uk}},
\author[A]{\fnms{Stuart} \snm{Barber}\corref{}\ead[label=e3]{stuart@maths.leeds.ac.uk}},
\author[C]{\fnms{Richard M.} \snm{Jackson}\ead[label=e4]{R.M.Jackson@leeds.ac.uk}}
\and
\author[D]{\fnms{Douglas L.} \snm{Theobald}\thanksref{t2}\ead[label=e5]{dtheobald@brandeis.edu}}
\runauthor{K. V. Mardia et al.}
\affiliation{University of Leeds and
University of Oxford, University of Nottingham,
University of Leeds, University of Leeds and Brandeis University}
\address[E]{K. V. Mardia\\
Department of Statistics\\
University of Oxford\\
Oxford, OX1 3TG\\
United Kingdom\\
\printead{e1}}
\address[A]{S. Barber\\
Department of Statistics \\
University of Leeds \\
Leeds, LS2 9JT \\
United Kingdom \\
\printead{e3}} 
\address[B]{C. J. Fallaize\\
School of Mathematical Sciences \\
University of Nottingham \\
Nottingham, NG7 2RD \\
United Kingdom \\
\printead{e2}\hspace*{7.7pt}}
\address[C]{R. M. Jackson\\
Institute of Molecular\\
\quad and Cellular Biology \\
University of Leeds \\
Leeds, LS2 9JT \\
United Kingdom \\
\printead{e4}}
\address[D]{D. L. Theobald\\
Department of Biochemistry \\
Brandeis University \\
415 South St \\
Waltham, Massachusetts 02454-9110 \\
USA \\
\printead{e5}}
\end{aug}

\thankstext{t2}{Supported by the US National Institutes of Health
Grants GM094468 and GM096053.}

\received{\smonth{2} \syear{2011}}
\revised{\smonth{11} \syear{2012}}

%
\begin{abstract}
We develop a Bayesian model for the alignment of two point
configurations under the full similarity transformations of rotation,
translation and scaling. Other work in this area has concentrated on
rigid body transformations, where scale information is preserved,
motivated by problems involving molecular data; this is known as form
analysis. We concentrate on a Bayesian formulation for statistical
\textit{shape} analysis. We generalize the model introduced by Green
and Mardia [\textit{Biometrika} \textbf{93} (2006) 235--254] for the
pairwise alignment of two unlabeled configurations to full similarity
transformations by introducing a scaling factor to the model. The
generalization is not straightforward, since the model needs to be
reformulated to give good performance when scaling is included. We
illustrate our method on the alignment of rat growth profiles and a
novel application to the alignment of protein domains. Here, scaling is
applied to secondary structure elements when comparing protein folds;
additionally, we find that one global scaling factor is not in general
sufficient to model these data and, hence, we develop a model in which
multiple scale factors can be included to handle different scalings of
shape components.
\end{abstract}

%

\end{frontmatter}

\section{Introduction}\label{secintro}

The shape of an object is the information about the object which is
invariant under the full similarity transformations
of rotation, translation and rescaling. In order to compare the shapes
of objects, we first seek to align them in some optimal registration.
In statistical shape analysis, objects often are reduced to a set of
points, known as landmarks, in $d$ dimensions and thus can
be represented as $m \times d$ point configurations, where $m$ is the
number of landmarks. Let $X$ be such a configuration of points; the
points on $X$ are $\mathbf{x}_j, j=1,\ldots,m$, where $\mathbf{x}_j \in
\mathbb{R}^d$ are the rows of $X$, with each row therefore giving the
coordinates of point $\mathbf{x}_j$. We shall consider the problem of
pairwise alignment, where the objective is to align one configuration,
such as $X$ above, with another configuration, $Y$, say, where the rows
of $Y$ are $\mathbf{y}_j, j=1,\ldots,m$, the locations of the points of $Y$.

Labeled shape analysis assumes a known, one-to-one correspondence
between the points on $X$ and $Y$, labeled such that $\mathbf{x}_j$
matches $\mathbf{y}_j, j=1,\ldots,m$. Since the configurations may be
presented in arbitrary registrations, it is necessary first to filter
out the similarity transformations so that only the shape information
of interest remains. Mathematically, the problem is to find $c$, $A$
and $\bolds{\tau}$ such that
\[
X^T = cAY^T + \bolds{\tau}\mathbf{1}_m^T,
\]
where $c > 0$ is a scaling parameter, $A$ is a $d \times d$ rotation
matrix and $\bolds{\tau} \in\mathbb{R}^d$ is a translation vector. Of
course, in practical situations, the point locations will be observed
with error, so the statistical problem is to find an optimal solution
to an equation of the form
\[
X^T = cAY^T + \bolds{\tau}\mathbf{1}_m^T
+ \Sigma,
\]
where $\Sigma$ is a $d \times m$ matrix of errors. The least squares
solution to this problem is the Procrustes solution [\citet{Shape}].

A much more challenging problem, which has been the subject of recent
research interest, is that of \textit{unlabeled} shape analysis, where
the correspondence between landmarks is not known and often the
configurations have different numbers of landmarks. Specifically, we
have an $m \times d$ configuration $X$ which we wish to align with an
$n \times d$ configuration $Y$, with $m \ne n$ in general. To keep
track of the correspondence between landmarks, an $m \times n$ matching
matrix $M = (m_{jk})$ is introduced, where $m_{jk} =1$ if $\mathbf{x}_j$ is
matched to $\mathbf{y}_k$ and $0$ otherwise. Thus, the problem is to
simultaneously estimate the matching matrix $M$ as well as to solve the
alignment problem described above for the labeled case. It is usually
assumed that any point on a configuration can match at most one point
on the other, so that any row or column of $M$ contains at most one
nonzero entry. Then the number of matched points, $0 \le L \le
\min(m,n)$, say, is $\sum_{j=1}^m \sum_{k=1}^n m_{jk}$ and is not
known. Thus, even for relatively small $m$ and $n$, the number of
possible matchings given by $M$ is large, which makes the problem very
challenging. Therefore, searching over all possible $M$ and optimizing
over transformation parameters to find a global solution is not
computationally feasible. \citet{GM} developed a Bayesian solution to
this problem, where the transformation, error and matching parameters
were all treated as unknown parameters and samples from the joint
posterior were drawn using MCMC. Although their model conceptually
could handle similarity transformations, their applications focused on
rigid-body alignment (i.e., no scaling parameter $c$)---inclusion
of the scaling parameter $c$ requires considerable attention, and it is
the purpose of the present paper to address this problem. In
particular, we reformulate the likelihood, which we find is necessary
for good performance, and derive the full conditional distribution for
the scaling parameter together with methods to sample from it.

Other solutions to the unlabeled alignment problem have been proposed.
One such method is to maximize a likelihood over the transformation
parameters conditional on a given matching, and then to propose a
different matching given the transformation parameters, and alternate
between these two steps; such methods have been used by \citet
{rodriguez10} and \citet{dryden07}. \citet{kent10} proposed a method
based on the EM algorithm, with the missing data being the labels
representing the matching between points. One problem with the methods
which alternate between matching and optimizing is that they can depend
on the initialization of the matching and can become trapped in local
modes [\citet{dryden07,kenobi2010b}]. \citet{schmidler07} developed a
fast matching method based on an approximation using geometric hashing,
and \citet{srivastava09} tackled the unlabeled problem by looking for
objects of predefined shape classes in cluttered point clouds, where
the points are samples from the outline of a shape.

An issue with methods in which the transformation parameters are
maximized out of the likelihood is that the alignment is considered
``correct,'' and uncertainty in this alignment is not fully propagated
[\citet{wilkinson2007}]. Therefore, it is desirable to consider a fully
Bayesian formulation, in which uncertainty in all the parameters is
correctly handled. Such a formulation for the case of unlabeled
similarity shape is the subject of this paper. \citet{TW} considered a
Bayesian model but concentrated on the labeled case and rigid-body
transformations. For more discussion on these points, and a deeper
comparison of the different methods, see, for example, the reviews by
\citet{green10} and \citet{mardia12}.

The paper is structured as follows: in Section \ref{secmodel} we
briefly review the model of \citet{GM} and introduce our generalization
of the model to full similarity transformations, with details of the
resulting conditional distribution for the scale factor $c$. We also
develop a model which can handle two scaling factors, which we find is
necessary to model the protein data in our applications. Section \ref
{secapplications} gives two illustrative examples: in the first we
consider the alignment of rat skulls, a data set which has been
analyzed previously in the shape analysis literature. In the second
example we introduce a novel application to the alignment of protein
domains based on a representation using their secondary structure
elements (beta strands and alpha helices). With this representation,
some scaling may allow for improved alignments between proteins which
have the same overall fold, but whose corresponding secondary structure
elements may be of different lengths; examples include homologous
proteins which have evolved from a common ancestor. We conclude the
paper with a discussion in Section \ref{secdiscussion}. Additional
results and material are provided in the \hyperref[app]{Appendix} and
in the supplementary material [\citet{mardia13}].

\section{The model}\label{secmodel}

Consider a pair of configurations of points in $d$ dimensions, $X$~and
$Y$, where $X$ consists of $m$ points and $Y$ of $n$ points. The
configurations $X$ and $Y$ can be represented by $m\times d$ and $n
\times d$ matrices, respectively, where the rows of $X$ are $\mathbf{x}_j
\in\mathbb{R}^d, j=1,\ldots, m$, and the rows of $Y$ are $\mathbf{y}_k \in
\mathbb{R}^d, k=1,\ldots, n$. In the model developed by \citet{GM} for
unlabeled landmarks,
\[
\mathbf{x}_j \sim N_d\bigl(\bolds{\mu}_{\psi_j},
\sigma^2 I_d\bigr),\qquad A\mathbf{y}_k
+ \bolds{\tau} \sim N_d\bigl(\bolds{\mu}_{\eta_k},
\sigma^2 I_d\bigr),
\]
where $\bolds{\mu}$ represents the (hidden) true point locations in some
space $V$ of volume~$v$, of which the observed configurations are noisy
realisations; the variance of the noise terms is $\sigma^2 I_d$. The
$\psi$ and $\eta$ are labels indexing the mapping between the
observed locations and $\bolds{\mu}$. In particular, if $\psi_j = \eta
_k$, then $\mathbf{x}_j$ and $\mathbf{y}_k$ are both generated by the same
point of $\bolds{\mu}$ and are therefore regarded as matched.
The mapping can be represented by a $m \times n$ matrix $M$ with
elements $m_{jk} = I(\psi_j = \eta_k)$, where $I(\cdot)$ is the
indicator function; $M$ is one of the parameters of interest about
which to draw inference. Each point on $X$ may be matched to at most
one point on $Y$ and vice versa. Therefore, each row and column of $M$
may contain at most one nonzero entry. Note that the case of labeled
landmarks is the special case where $M$ is known.

\subsection{Likelihood}\label{sseclik}

For our full similarity transformation model, we consider a different
formulation to that of \citet{GM}. Rather than considering one
configuration being transformed into the space of the other, we
initially consider a more ``symmetrical'' formulation where both
configurations are transformed into $\bolds{\mu}$-space, which can be
thought of as an ``average space.'' Specifically, we have
%
\begin{equation}
\label{eqscalemodel}\quad \frac{1}{\sqrt{c}}B^{T} \mathbf{x}_j +
\bolds{\tau}_1 \sim N_d\bigl(\bolds{\mu}_{\psi_j},
\sigma^2 I_d\bigr),\qquad \sqrt{c}B
\mathbf{y}_k + \bolds{\tau}_2 \sim N_d\bigl(\bolds{
\mu}_{\eta_k},\sigma^2 I_d\bigr),
\end{equation}
where $c>0$ is a scale parameter, $B$ is a $d \times d$ rotation matrix
and $\bolds{\tau}_1, \bolds{\tau}_2 \in\mathbb{R}^d$
are translation vectors; $B^T$ denotes the transpose of $B$. We have
\[
\frac{1}{\sqrt{c}}B^{T}\mathbf{x}_j + \bolds{
\tau}_1 = \bolds{\mu}_{\xi
_j} + \bolds{\varepsilon}_{1j},\qquad
j=1,\ldots,m,
\]
and
\[
\sqrt{c}B\mathbf{y}_k + \bolds{\tau}_2 = \bolds{
\mu}_{\eta_k} + \bolds{\varepsilon}_{2k},\qquad k=1,\ldots,n,
\]
where the $\bolds{\varepsilon}$ represent errors in the observed point
locations. Assuming Gaussian errors, so $\bolds{\varepsilon}_{1j}, \bolds
{\varepsilon}_{2k} \sim N(0,\sigma^2 I_d)$, results in model (\ref
{eqscalemodel}). (We note that other error structures, such as
heavy-tailed distributions, could be used, and this is computationally
feasible. This would allow for the possibility of outliers. However,
this would have the undesirable effect of including matches which are
far apart after transformation, so the standard notion of robustness is
not meaningful for alignment.) We denote the density of the error terms
by $ f(\bolds{\varepsilon}) = \phi(\bolds{\varepsilon}/\sigma)/\sigma^d$,
where $\phi(\cdot)$ is the standard normal distribution in $d$
dimensions. We now derive the full likelihood of the observed data. The
points on $\bolds{\mu}$ are regarded as uniformly distributed over the
region $V$. Assuming boundary effects can be ignored, then the
likelihood contribution of the unmatched $X$ points is therefore
\[
\prod_{j:m_{jk}=0\ \forall k} c^{-d/2} \frac{1}{v} \int
_V f \biggl(\frac{1}{\sqrt{c}}B^{T}
\mathbf{x}_j + \bolds{\tau}_1 - \bolds{\mu} \biggr)\,d\bolds{\mu} =
v^{-(m-L)} c^{-d(m-L)/2}.
\]
Similarly, the contribution of the unmatched $Y$ points is
\[
\prod_{k:m_{jk}=0\ \forall j} c^{d/2} \frac{1}{v} \int
_V f(\sqrt{c}B\mathbf{y}_k + \bolds{
\tau}_2 - \bolds{\mu})\,d\bolds{\mu} = v^{-(n-L)} c^{d(n-L)/2},
\]
and the contribution of the matched points between $X$ and $Y$ is
\[
\prod_{j,k:m_{jk}=1} c^{-d/2} c^{d/2}
\frac{1}{v} \int_V f \biggl(\frac{1}{\sqrt{c}}B^{T}
\mathbf{x}_j + \bolds{\tau}_1 - \bolds{\mu} \biggr) f(\sqrt{c}B
\mathbf{y}_k + \bolds{\tau}_2 - \bolds{\mu})\,d\bolds{\mu}.
\]
We have
\[
\int_V f(z+u)f(u)\,du = g(z),
\]
the density of $\bolds{\varepsilon}_{1j} - \bolds{\varepsilon}_{2k}$. Here, $ z
= \frac{1}{\sqrt{c}}B^{T}\mathbf{x}_j + \bolds{\tau}_1 - \sqrt{c}B\mathbf
{y}_k - \bolds{\tau}_2$ and $g(z) = \phi(z/\sqrt{2}\sigma)/(\sqrt
{2}\sigma)^d $. The complete likelihood, $p(x,y;M,B,\bolds{\tau}_1,\bolds
{\tau}_2,\sigma,c)$, is then
\[
v^{-(m+n)+L} c^{{d(n-m)}/{2}} \times\prod_{j,k:m_{jk}=1}
\frac{\phi\{(B^{T}\mathbf{x}_j/{\sqrt{c}} + \bolds
{\tau}_1 - \sqrt{c}B\mathbf{y}_k - \bolds{\tau}_2)/(\sqrt{2}\sigma)\}
}{(\sqrt{2}\sigma)^d}.
\]
Also, $p(M) \propto(\frac{\kappa}{v})^L $, which results from a
model in which the unobserved $\bolds{\mu}$ points are realizations of a
homogeneous Poisson process over the region $V$ [\citet{GM}]. This
process is thinned so that each $\bolds{\mu}$ point generates an
observed point of exactly one of the following forms: on $X$ only, on
$Y$ only, on both $X$ and $Y$, or not observed. The $\bolds{\mu}$ points
generating an observation on $X$ and $Y$ are the matched points. The
probabilities of the thinned process are parameterized by~$\kappa$,
which can be regarded as the propensity of points to be matched a
priori. In particular, larger values of $\kappa$ give a stronger prior
preference to larger numbers of matched points.

Combining these terms, the joint model $p(M,B,\bolds{\tau}_1,\bolds{\tau
}_2,\sigma,c,x,y)$ is proportional to
\begin{eqnarray*}
&& p(B)p(\tau_1)p(\tau_2)p(c)p(\sigma)c^{{d(n-m)}/{2}}
\bigl(\sigma^2\bigr)^{-Ld/2} \kappa^L
\\
&&\qquad{}\times \exp\biggl\{-\frac{1}{4\sigma^2} \sum_{j,k:m_{jk}=1}\biggl\|
\frac{1}{\sqrt{c}}B^{T}x_j + \tau_1 - \sqrt
{c}By_k - \tau_2 \biggr\|^2 \biggr\}.
\end{eqnarray*}
This can be written as
%
\begin{eqnarray}
\label{eqscalejointmodel}\quad p(M,A,\bolds{\tau},\sigma_c,c,x,y) &\propto&
p(A)p(\bolds{\tau})p(c)p(\sigma_c)c^{{d}(n-m+L)/{2}}\bigl(
\sigma_c^2\bigr)^{-Ld/2}\kappa^L
\nonumber\\[-8pt]\\[-8pt]
&&{} \times\exp\biggl\{-\frac{1}{4\sigma_c^2} \sum_{j,k:m_{jk}=1}
\|\mathbf{x}_j - cA\mathbf{y}_k - \bolds{\tau} \|^2
\biggr\},
\nonumber
\end{eqnarray}
where $A = B^2$, $ \bolds{\tau} = \sqrt{c}B(\bolds{\tau}_2-\bolds{\tau
}_1) $ and $\sigma_c^2 = c\sigma^2$. The parameter $\sigma_c^2 $ can
be regarded as the variance of the errors in $X$-space, and the term in
the exponent above is of the same form as the transformation of the $Y$
points into $X$-space as in \citet{GM}, with the scaling parameter $c$
now included. Note that the exponent of $c$ is now $\frac
{d(n-m+L)}{2}$ as opposed to $nd$, as would result from strictly
following the original formulation in \citet{GM}; we find our novel
formulation provides much better performance when dealing with full
similarity shape. [Note that, although \citet{GM} provided a general
formulation which could deal with similarity transformations, they
focused on rigid body transformations only in their practical
applications; the implementation was not considered.] Intuitively, it
is plausible to expect that the exponent of $c$ should depend on the
number of matched points $L$, rather than the fixed quantity $nd$, and
that is the case with our formulation; this is a possible explanation
for the improved performance.

\subsection{Prior distributions and MCMC updates}\label{ssecsccond}

Priors for the parameters $A$, $\bolds{\tau}$, $\sigma_c$ and $M$ are
of the same form as in \citet{GM}. The rotation matrix $A$ has a
matrix-Fisher prior distribution, where $p(A)
\propto\exp\{\operatorname{tr}(F_0^TA) \}$ and the parameter $F_0$ is a
$d \times d$ matrix. $A$ is parameterized by one angle $\theta$ when
$d=2$, and by Eulerian angles, $\theta_{12},\theta_{13},\theta_{23}$,
say, in the case $d=3$. In our examples of Sections \ref{secrat} and
\ref {secproteins}, we use a uniform prior on $A$, which is the special
case where $F_0$ is the $d \times d$ matrix of zeroes. $A$ then has a
uniform prior with respect to the invariant measure on $SO(3)$, the
Haar measure, where $SO(3)$ is the special orthogonal group of all $d
\times d$ rotation matrices. With our parameterization, this measure is
$\cos(\theta_{13})\,d\theta_{12}\,d\theta_{13}\,d\theta_{23}$. For the
translation vector $\bolds{\tau}$, we have $\bolds{\tau} \sim
N_d(\bolds {\mu}_{\tau},\sigma_{\tau}^2 I_d)$, where
$\bolds{\mu}_{\tau}$ is a mean vector and $\sigma_{\tau}^2 I_d$ a
covariance matrix, with $I_d$ the $d \times d$ identity matrix. For the
noise parameter $\sigma_c$, we have $\sigma_c^{-2}
\sim\Gamma(\alpha,\beta) $, where $p(\sigma _c^{-2})
\propto\sigma_c^{-2(\alpha-1)}\exp(-\frac{\beta }{\sigma_c^{2}} ) $.
The matching matrix $M$ is parameterized by $\kappa> 0$, with $p(M)
\propto(\frac{\kappa}{v})^L $ as described above.

We perform inference by generating samples from the posterior
distribution (\ref{eqscalejointmodel}) using MCMC. Updates for the
parameters $A$, $\bolds{\tau}$, $\sigma$ and $M$ take the same form as
in \citet{GM}, with the necessary adjustments being made to the various
terms to include the scale factor $c$ where appropriate. We now
concentrate on the scale parameter $c$.

From (\ref{eqscalejointmodel}), the conditional distribution of $c$
is proportional to
%
\begin{equation}
\label{eqcconditional} p(c)c^{{(n-m+L)d}/{2}}\exp\biggl(-\frac
{1}{4\sigma_c^2}\sum
_{j,k:m_{jk}=1}\|\mathbf{x}_j - cA
\mathbf{y}_k -\bolds{\tau}\|^2 \biggr),
\end{equation}
where $L = \sum_{j=1}^{m}\sum_{k=1}^{n} m_{jk} $ is the number of
matched points. Adopting a gamma prior on $c$ with parameters $\alpha
_c$ and $\lambda_c$, so that $p(c) \propto c^{\alpha_c -1} \exp
(-\lambda_c c)$, we have the conditional distribution
%
\begin{equation}
\label{eqcconditional2} p(c|A,\bolds{\tau},\sigma_c,M,X,Y) \propto
c^{r -1} \exp\bigl(-\tfrac
{1}{2}\nu c^2 +\delta c
\bigr),
\end{equation}
where $r=\frac{(n-m+L)d}{2} + \alpha_c $ and
\[
\nu= \sum_{j,k:m_{jk}=1} \mathbf{y}_k^T
\mathbf{y}_k/2\sigma_c^2,\qquad \delta=
\sum_{j,k:m_{jk}=1}(\mathbf{x}_j - \bolds{
\tau})^TA\mathbf{y}_k/2\sigma_c^{2}
- \lambda_c.
\]
This distribution is a member of the generalized exponential family of
distributions introduced by \citet{lye93}. In particular, it is in the
form of the generalized gamma distribution of \citet{creedy94}, who
used these distributions for modeling the stationary distribution of
prices in economic models. This generalized gamma distribution has the form
%
\begin{equation}
\label{eqgengam} p(c) = \exp\bigl(\zeta_1 \log c +
\zeta_2 c + \zeta_3 c^2 + \zeta_4
c^3 -\eta\bigr)
\end{equation}
for $c > 0$, where $\zeta_i, i=1,\ldots,4$ are
parameters and $\eta$ is a constant. Here we have the special case
$\zeta_4=0$ in (\ref{eqgengam}), which we shall denote as the
halfnormal-gamma distribution. We are not aware of other work which
considers this particular distribution or methods to simulate from it.
We use a Metropolis step, and also devise an acceptance-rejection
algorithm, details of which are in the supplementary material [\citet
{mardia13}]. Note that the choice of a gamma prior led to conjugacy,
since both the likelihood term in (\ref{eqcconditional}) and the
conditional posterior for $c$ are of halfnormal-gamma form; therefore,
our acceptance-rejection method can be used to generate exact samples
from this full conditional distribution.
In our applications, we have used the Metropolis method to perform
updates for $c$, which we give details of here. A proposal value $c'$
is generated, given the current value $c$, from the distribution
\[
c'|c \sim N\bigl(c,w^2\bigr),
\]
where
\[
w = \bigl\{\nu+ (r-1)/s_m^2 \bigr\}^{-{1}/{2}}
\]
and $s_m = \{\delta+ \sqrt{\delta^2 +4(r-1)\nu}\}/2 \nu$ is the
mode of the conditional distribution (\ref{eqcconditional2}). The
acceptance probability for the Metropolis step is
\[
\alpha_p = \min\bigl[1,\bigl(c'/c
\bigr)^{{(n-m+L)d}/{2}+\alpha_c
-1}\exp\bigl\{-\tfrac{1}{2}\nu\bigl(c'^2-c^2
\bigr) + \delta\bigl(c'-c\bigr) \bigr\} \bigr].
\]
The choice of proposal distribution is motivated by a general principle
of normal approximations to members of the exponential family of
distributions. Details are given in the \hyperref[app]{Appendix}, where, in particular,
we show $w^2$ to be an approximate variance for the halfnormal-gamma
conditional distribution of $c$ under such a normal approximation. The
success of the Metropolis method will depend on how well the proposal
distribution approximates the target distribution. Hence, in situations
where this normal approximation is less adequate, the
acceptance-rejection method may be more efficient. However, we find the
Metropolis method is perfectly adequate for our examples (where the
configurations have relatively small numbers of points) and as such
is used throughout.

\subsection{The two-scale model}

We now develop a model which allows for more than one scaling
parameter, motivated by our protein folding applications in Section
\ref{secproteins}. Suppose there are two sets of points, groups $0$
and $1$, say, with the points in each group subject to different
transformations. We assume that matched points, where $m_{jk}=1$, are
from the same group. Introduce\vspace*{1pt} class labels $z_{j}^{x} \in\{0,1\},
j=1,\ldots,m$, to denote the group of point $\mathbf{x}_j$, and
likewise for the $Y$ points. For group $0$ we have
\[
\frac{1}{\sqrt{c_0}}B_0^{T}\mathbf{x}_j + \bolds{
\tau}_1^0 = \bolds{\mu}_{\xi_j} + \bolds{
\varepsilon}_{1j},\qquad j=1,\ldots,m,
\]
and
\[
\sqrt{c_0}B_0\mathbf{y}_k + \bolds{
\tau}_2^0 = \bolds{\mu}_{\eta_k} + \bolds{
\varepsilon}_{1k},\qquad k=1,\ldots,n,
\]
and similarly for the group $1$ points. Let $m_0$ and $n_0$ denote the
number of $X$ and $Y$ points, respectively, in group $0$, and similarly
for $m_1$ and $n_1$. Also let $L_0$ and $L_1$ be the number of matched
points in group $0$ and $1$, respectively. Using similar arguments to
those in Section \ref{sseclik}, the joint model $p(M,A,\bolds{\tau
}^{0},\bolds{\tau}^{1},\sigma_{c_0},\sigma_{c_1},c_0,c_1,\mathbf{x},\mathbf
{y})$ is proportional to
\begin{eqnarray*}
&&
p(A)p\bigl(\bolds{\tau}^{0}\bigr)p\bigl(\bolds{\tau}^{1}
\bigr)p(c_0)p(c_1)p(\sigma_{c_0})p(
\sigma_{c_1})\kappa^L
\\
&&\qquad{}\times \bigl(\sigma_{c_0}^2\bigr)^{-L_0d/2}
c_0^{{d}(n_0-m_0+L_0)/{2}} \\
&&\qquad{}\times \exp\biggl\{-\frac{1}{4\sigma_{c_0}^2}
\sum
_{j,k:m_{jk}=1,z_{j}^{x}=0} \bigl\|\mathbf{x}_j - c_0A
\mathbf{y}_k - \bolds{\tau}^{0} \bigr\|^2 \biggr\}
\\
&&\qquad{}\times \bigl(\sigma_{c_1}^2\bigr)^{-L_1d/2}
c_1^{{d}(n_1-m_1+L_1)/{2}} \\
&&\qquad{}\times \exp\biggl\{-\frac{1}{4\sigma_{c_1}^2}
\sum
_{j,k:m_{jk}=1,z_{j}^{x}=1} \bigl\|\mathbf{x}_j - c_1A
\mathbf{y}_k - \bolds{\tau}^{1} \bigr\|^2 \biggr\},
\end{eqnarray*}
where $ \bolds{\tau}^{0} = \sqrt{c_0}B_0(\bolds{\tau}^0_2-\bolds{\tau
}_1^0) $ and $\sigma_{c_0}^2 = c_0\sigma^2$, and likewise for group
$1$ parameters. We assume that both groups have the same rotation
$A=B_0^2=B_1^2$; if there is no translation (as in our protein
applications in Section \ref{secproteins}), this assumption
corresponds to a model where the entire configurations are first
rotated by $A$, before the appropriate scaling is applied to each
individual element. This is exactly the behavior we require in the
protein alignment applications of Section \ref{secproteins} when
using our representation of protein secondary structure.
Assuming that the priors for the scale and noise parameters are
independent and have the same form as previously, then the updates for
the Markov chain have the same form, with the relevant updates for the
transformation parameters for each group naturally depending on only
the points in that group. Additionally, we also propose a switch of the
class labels at each iteration of the chain. For identifiability of the
groups, we set $c_1 > c_0$.

\section{Applications}\label{secapplications}

\subsection{Rat growth (labeled landmarks)}\label{secrat}

In growth data, the interest is to assess changes in shape over time.
Here, size is a key concept, since growth leads to an increase in the
object's overall size, while its shape may remain the same. Hence,
scaling information is highly relevant and must be taken into account
during the alignment process.

We illustrate our method on data relating to the growth of a rat's
skull. The data are described in Bookstein [(\citeyear{bookstein91}), page 67] and the
references therein, and have been analyzed by many other authors
including \citet{kent01b}, \citet{kent01a} and \citet{kenobi10}. The
data consist of $m = 8$ landmark locations in $d=2$ dimensions on the
skulls of $21$ laboratory rats measured at $8$ timepoints between the
ages of $7$ and $150$ days. The correspondence between landmarks is
known, hence, this is an example of labeled shape analysis. Since the
measurements are on the same rat at different ages, we would expect
clear differences in the size of the rat and, hence, there may be a
change in scale, as well as possible changes in shape. The real
interest is in changes in shape over time, but the configurations from
each timepoint must first be registered by removing the information not
relating to shape. Since the rat will grow over time, it is necessary
to remove size information and, hence, the full similarity
transformations must be used in the registration.

\begin{figure}

\includegraphics{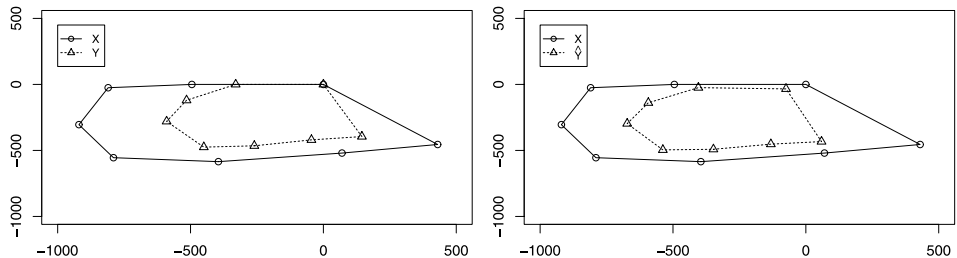}

\caption{Pairwise alignment between the rat configurations at
timepoints $1$ and $8$ using only rotation and translation. The left
panel shows the unregistered configurations, and the right panel the
registered configurations; the need for scaling is clearly
apparent.}\label{figratgrowth2}
\end{figure}

\begin{figure}

\includegraphics{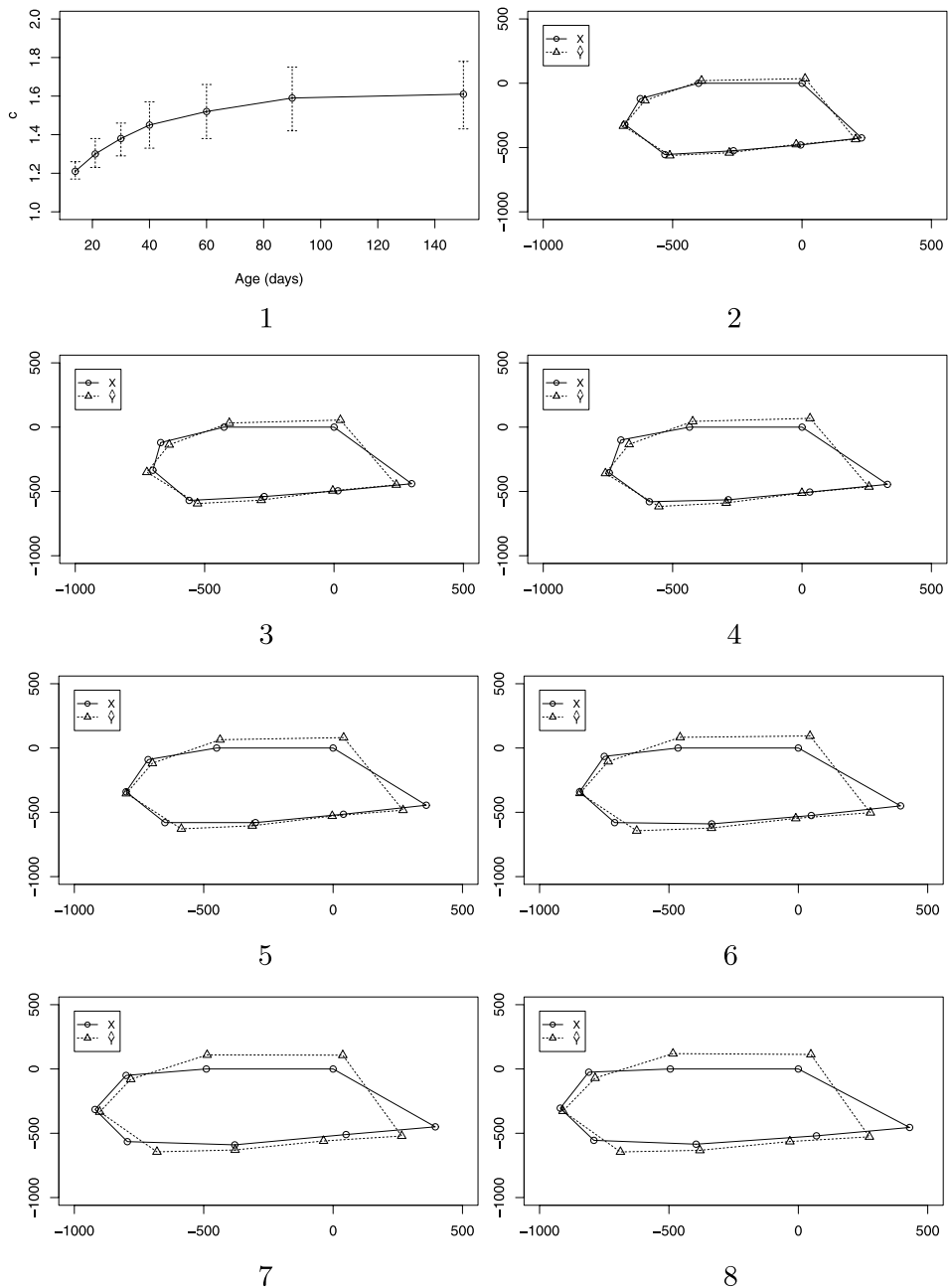}

\caption{Pairwise alignments between the rat configurations at
timepoints $2$ to $8$ and timepoint~$1$. In panel $1$, the posterior
medians for the scale factor $c$ are plotted against the age of the
rat, together with $95 \%$ posterior intervals. Panels $2$ to $8$ show
the corresponding superpositions, ordered chronologically, with panel
$2$ showing the alignment between timepoints $2$ and $1$ and so on; in
each case the dashed line represents the estimated superposition of the
skull at the first timepoint onto the skull at the later timepoint
(solid line).}\label{figratgrowth}
\end{figure}

To highlight the need to include scaling in the alignment, we first
consider using only a rigid-body transformation. In Figure \ref
{figratgrowth2} we see the initial configurations of the rat at the
first and last time point, and the registered configurations using only
rotation and translation as in the original method of \citet{GM}. Here,
the need for scaling is evident when comparing the fit to that obtained
by using the full similarity transformation (Figure \ref
{figratgrowth}, bottom right).

We illustrate our method of full similarity shape alignment on one rat
[labeled~$1$ in \citet{bookstein91}] by comparing the shape at
timepoints $2$ to $8$ with the shape at the first timepoint. We set a
diffuse prior on the translation $\bolds{\tau}$, with $\bolds{\mu}_{\tau
}$ equal to the difference in centroids and $\sigma_{\tau} =1000$. We
use an exponential prior for $c$ with mean $1$, and set $\alpha=1$ and
$\beta=8$, but find that the results are robust to these settings for
$\alpha$ and $\beta$. In each case, we denote the younger rat
configuration by $Y$ and the older one by $X$.

Panel $1$ of Figure \ref{figratgrowth} shows the posterior median of
the scale factor $c$ from each of the seven pairwise alignments of the
youngest rat configuration with the older ones, together with a $95 \%$
posterior interval. Here we clearly see an initially rapid increase in
the scale factor, slowing as the rat gets older. Panels $2$ to $8$ show
the corresponding superpositions of the younger rat configuration
($\hat{Y}$) onto the older one ($X$), with the transformation obtained
using the posterior mean estimates of $A$, $\bolds{\tau}$ and $c$. As
well as an increase in size, there is also evidence of a change in
shape, as seen by the progressively looser fits as the rat gets older.
In particular, the skull becomes longer and thinner as the rat gets older.

\subsection{Aligning protein domains}\label{secproteins}

\subsubsection{Proteins and secondary structure}\label{subsecproteinSS}

We now consider an application to the alignment of protein domains. A
protein is a chain of amino acid residues, and there are $20$ different
amino acid types. An amino acid consists of a structure common to all
amino acid types, plus an additional side-chain structure which
determines which of the $20$ types it is. In particular, every amino
acid contains an alpha-carbon ($C_{\alpha}$) atom, and one possible
representation of protein shape is the configuration of $C_{\alpha}$
atoms. Indeed, the first statistical work involving 3-d protein data in
bioinformatics began with \citet{wu}, who used the alpha-carbon
($C_{\alpha}$) atom of each amino acid residue as a landmark location.

\begin{figure}

\includegraphics{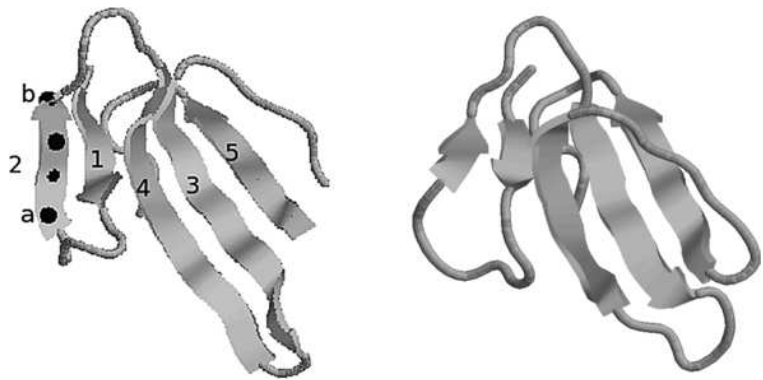}

\caption{Left: the domain 2VLWA00. The domain consists of $5$ beta
stands, labeled $1$--$5$ to denote their sequence order in the chain.
The locations of the $C_{\alpha}$ atoms (the dark spheres) from strand
$2$ are shown for illustration; this strand has $4$ residues and hence
$4$ $C_{\alpha}$ atoms. Right: the domain 1FASA00.}\label{fig2VLWsse}
\end{figure}

We use a representation based on the secondary structure elements of a
protein. At the secondary structure level, a protein can be represented
in terms of $\beta$ strands and $\alpha$ helices (the two main
secondary structure elements), which are themselves sub-chains of amino
acid residues. The spatial arrangement of these elements, together with
their connectivity, determine the fold of the protein, which is crucial
to the biological activity of the protein. An example is shown in the
left panel of Figure \ref{fig2VLWsse}, the domain 2VLWA00 which we
use in our examples below. The arrows represent beta strands, which are
labeled $1$--$5$ to indicate the sequence order in which they appear in
the chain, each made up of a number of amino acid residues. For
illustration, we have shown the positions of the $C_{\alpha}$ atoms of
each residue on the strand labeled $2$, represented by the dark spheres
(not to scale); this particular strand has $4$ residues and hence $4$
$C_{\alpha}$ atoms. The beta strands are joined together by further
regions of the amino acid chain, known as loops, represented here by
the thin strings. For a thorough introduction to protein secondary
structure, see, for example, \citet{Tooze}, Chapter~2.

One possible approach is to represent an element by the centroid of the
$C_{\alpha}$ atoms from the residues of the element. The protein would
then be reduced to a configuration of points, with each point
representing the centroid of one element. However, applying scaling to
this representation would also scale the distances between secondary
structure elements in the packing arrangement of the protein. This is
not satisfactory since these distances should be preserved. Where
scaling is really required is in the comparison of the lengths of the
elements. Two proteins sharing the same fold may have a very similar
spatial arrangement of secondary structure elements, but the lengths of
the elements may be longer in one than the other. Hence, we consider an
approach using vectors to represent secondary structure elements, where
each distinct element is represented by a vector from the origin to a
single point. Scaling a configuration then only alters the length of
each vector. The vector representation is found by taking the
difference between the start and end points of the principal axis
through the element, found using the method described in \citet{Taylor}
as follows. For any particular element (a beta strand say), the
locations of the $C_{\alpha}$ atoms are taken to form a data cloud in
three dimensions, and the principal axis (essentially the first
principal component) is calculated. The start and end points of each
element are then found by orthogonally projecting the $C_{\alpha}$
atom of the first and last residues onto this axis; the difference
between these two points is then the point (vector) representing the
element. For example, relating to Figure \ref{fig2VLWsse}, to find
the point representing the strand labeled $2$, the principal axis
through the $4$ $C_{\alpha}$ atoms is first found. The start point of
the strand is then found by projecting the $C_{\alpha}$ atom from the
first residue of the element (labeled a) onto this axis; similarly, the
end point of the strand is found by projecting the $C_{\alpha}$ atom
from the last residue of the element (labeled b) onto the axis. The
difference between the end and start points is then the vector
representing this strand. In this example, there are $5$ strands and,
hence, there will be $5$ points in total representing the protein, each
found using the procedure above. Notationally, we represent the start
point of the $j$th element of one protein by $\mathbf{x}_{1j}$ and
the end point by $\mathbf{x}_{2j}$. The $j$th row of configuration
$X$ is then $ \mathbf{x}_{j} = \mathbf{x}_{2j} - \mathbf{x}_{1j}$, with a
similar definition for the $k$th row of the second protein, $Y$,
say, $\mathbf{y}_{k}$.

To address the particular challenges faced in the protein examples, we
make two alterations to the methodology used in the paper thus far.
First, the ordering of the secondary structure elements is important,
as proteins which evolve from a common ancestor do so via the
insertion/deletion of amino acid residues, and ultimately possibly
secondary structure elements. As such, the parts which are
conserved/common between two proteins will be placed in the same order
relative to each other. Hence, we only allow proposed updates to the
matching matrix $M$ which preserve the sequence order of the elements.
Second, in contrast to traditional applications in shape analysis,
there is no reason why we should expect a single global scaling factor
to be appropriate, since different pairs of secondary structure
elements may require different scaling. Therefore, we propose a model
with two scaling factors, which are sufficient to provide a good fit to
the data in our examples, as each configuration has only a relatively
small number of points (a~maximum of ten). This model could be readily
extended to handle a general number of scalings, which may be required
for configurations with a larger number of points.

\subsubsection{Illustrative examples}

We illustrate this approach using $3$ protein domains each consisting
of beta strands: 2VLWA00, 1FASA00 and 1M9ZA00, which are classified in
the same superfamily (CATH code 2.10.60.10) in the CATH database [\citet
{CATH}]; the domain names refer to their respective CATH identification
labels. Since they are classified in the same CATH superfamily, they
have the same fold and, hence, the domains should possess a high degree
of structural similarity. However, the individual strands will not
necessarily have the same length, so some scaling may be required to
produce a good alignment of the individual structural elements
(points). We provide two examples, namely, the domain 2VLWA00 aligned
with each of the domains 1FASA00 and 1M9ZA00.

Our first example is the pair of domains 2VLWA00 (configuration $X$)
and 1FASA00 (configuration $Y$), each of which consists of five beta
strands; the structures are shown in Figure \ref{fig2VLWsse}. We used
the settings $\alpha= \beta=1$ throughout this section for the prior
of $\sigma^{-2}$. For the scale factor $c$, we have an exponential
prior with the mean parameter taken as $1$ ($\alpha_c = 5$, $\lambda
_c = 5$) and we use $\kappa= 100\mbox{,}000$. We do not allow for translation,
since translation is removed when taking the difference between start
and end points of an element. The matches obtained and their respective
probabilities are given in Table \ref{tabbetasse1}. We see that each
pair of points matches with high probability. The posterior median of
$c_0$ is $1.06$, with $95$\% posterior interval $(0.75,1.56)$, and the
posterior median of $c_1$ is $1.64$, with $95$\% posterior interval
$(1.38,2.14)$. These results highlight the ability of the model to
capture the different scaling required for different elements, which we
now explore further.

\begin{table}
\caption{Matches from an alignment of the secondary structures of 2VLW
($X$) and 1FAS ($Y$)}\label{tabbetasse1}
\begin{tabular*}{\tablewidth}{@{\extracolsep{\fill}}lccccccc@{}}
\hline
\textbf{Match} & $\bolds{x}$ & $\bolds{y}$ & \textbf{prob}
& $\bolds{\|x\|/\|y\|}$ & \textbf{prob (no scale)} & \textbf{prob (global
scale)} & $\bolds{\hat{f}_0}$ \\
\hline
1 & 1 & 1 & 0.989 & 2.41 & 0.869 & 0.983 & 0.17 \\
2 & 2 & 2 & 0.945 & 2.70 & 0.701 & 0.958 & 0.15 \\
3 & 3 & 3 & 0.968 & 1.62 & 0.347 & 0.965 & 0.08 \\
4 & 4 & 4 & 0.980 & 1.59 & 0.414 & 0.970 & 0.10 \\
5 & 5 & 5 & 0.924 & 1.02 & 0.672 & 0.512 & 0.98 \\
\hline
\end{tabular*}
\end{table}

Column $5$ of Table \ref{tabbetasse1} shows the length ratios of the
matched points prior to any scaling. This suggests that some scaling is
certainly necessary, and further still that varying amounts of scaling
may be necessary for different pairs of points to provide the best fit
to the data. We now consider the improvement in fit offered by first
introducing one scaling factor, and the further improvement offered by
adding a second scaling factor. The matches obtained using no scaling
and one global scale factor are shown in columns $6$ and $7$ of Table
\ref{tabbetasse1}, respectively. For the case of one global scale
factor, where the posterior median of $c$ is $1.54$, with $95$\%
posterior interval $(1.25,1.85)$, the model is clearly not sufficient
to capture all the matches with high probability. In particular, the
match between $\mathbf{x}_5$ and $\mathbf{y}_5$ has a much lower posterior
probability of $0.512$; this can be explained due to the ratio of
lengths being $1.02$, in comparison to the other ratios, which are
$1.59$ and above. However, the inclusion of a scaling parameter offers
a clear improvement over the case where no scaling is applied
whatsoever, as seen by the substantially lower matching probabilities
obtained when no scaling is used.

Looking purely at the length ratios of the matched points, one might
consider whether a model with three groups might be necessary. Column
$8$ of Table \ref{tabbetasse1} shows the empirical proportion of the
iterations that each pair of matched points were in group $0$ (the
group with the smaller scale factor), $\hat{f}_0$ say. These
proportions suggest that the points are separated into two clear
groups, with the match between $\mathbf{x}_5$ and $\mathbf{y}_5$ being
accounted for in its own group, and that group $1$ can accommodate the
other matches; this evidence, together with the posterior
probabilities, suggests that two scaling factors are sufficient in this
case. The model could readily accommodate more scaling factors, but
given the small number of points in this example, this appears
excessive and risks overfitting.

\begin{figure}

\includegraphics{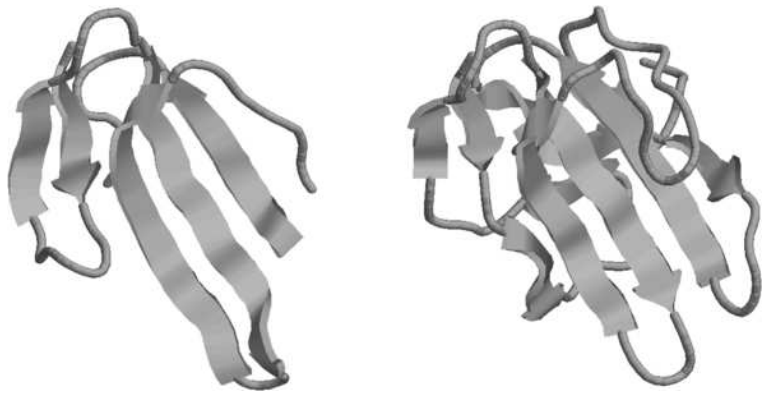}

\caption{Left: the domain 2VLWA00, which has $5$ beta strands, from
the first example. Right: the domain 1M9ZA00, which has $10$ beta
strands.}\label{fig2VLW1M9Zsse}
\end{figure}

\begin{table}
\caption{Matches from an alignment of the secondary structures of 2VLW
($X$) and 1M9Z ($Z$), for the cases of two scaling factors and one
global scaling factor} \label{tabbetasse3}
\begin{tabular*}{\tablewidth}{@{\extracolsep{\fill}}lcccccc@{}}
\hline
\textbf{Match} & $\bolds{x}$ & $\bolds{z}$ & \textbf{prob (two scale)}
& \textbf{prob (global scale)} &
$\bolds{\|x\|/\|z\|}$ & $\bolds{\hat{f}_0}$ \\
\hline
1 & 1 & 1 & 0.990 & 0.975 & 1.29 & 0.07 \\
2 & 2 & 4 & 0.988 & 0.980 & 1.11 & 0.04 \\
3 & 3 & 5 & 0.997 & 0.985 & 1.47 & 0.04 \\
4 & 4 & 6 & 0.990 & 0.981 & 1.19 & 0.06 \\
5 & 5 & 9 & 0.957 & 0.179 & 0.81 & 0.99 \\
\hline
\end{tabular*}
\end{table}

To illustrate further the power of the unlabeled method, we now
consider an example with an unequal number of points (secondary
structure elements). The domains are 2VLWA00 from the previous example
(configuration $X$) which has five beta strands, and 1M9ZA00
(configuration $Z$) which has ten beta strands; the structures are
shown in Figure \ref{fig2VLW1M9Zsse}. Even with the full possible
matching of five points, the matching between individual strands is not
obvious. (In the previous example, there is only one possible matching
matrix $M$ consistent with five matches, due to the ordering
constraint.) The posterior matches in this example, for the cases of
two scaling factors and one global scaling factor, are shown in Table
\ref{tabbetasse3}. For the first case, five matches are found with
high posterior probabilities. The empirical proportion of the
iterations each match spent in group $0$ is again shown, in column $7$
of Table \ref{tabbetasse3}. As in the first example, the model
clearly separates the matches into two groups, with the match requiring
a smaller scaling factor being accommodated in group $0$; the matching
probabilities for the global scale model show that this match is
neglected when only one scaling factor is used. We note that the beta
strand represented by $\mathbf{x}_5$ is in group $0$ in both cases; this
strand is consistently smaller in domain 2VLWA00 than in the other
domains we have considered. This evidence again suggests that one
global scale factor is not sufficient to capture all possible matches
with high probability, but that a two-scale model is adequate.

\subsection{Sensitivity to prior settings and computational issues}

The role of and sensitivity to the parameters $\kappa$ and $\beta$
were discussed in \citet{GM}; higher values of $\kappa$ encourage more
matches, and $\beta$ is an inverse scale parameter for the noise
variance, with larger values leading to generally fewer matches. Here,
we concentrate on the sensitivity of our results to the prior settings
for the scaling parameters. We consider three cases, namely, $\alpha_c
= 0.1,5.0,10.0$; in each case, we set $\lambda_c = \alpha_c$, giving
a prior mean of $1$, with larger values of $\alpha_c$ corresponding to
a smaller prior variance. The matches obtained for our first protein
example, the pair 2VLW-1FAS, are shown in Table \ref{tabbetasse4}.
The effect on the parameters $c_0$ and $c_1$ can be seen in Table \ref
{tabbetasse5}.

\begin{table}
\tabcolsep=4pt
\caption{Matches from an alignment of the secondary structures of 2VLW
($X$) and 1FAS ($Y$) for three different prior settings}
\label{tabbetasse4}
\begin{tabular*}{\tablewidth}{@{\extracolsep{\fill}}lccccc@{}}
\hline
\textbf{Match} & $\bolds{x}$ & $\bolds{y}$ &
\multicolumn{1}{c}{\textbf{prob (case 1:} $\bolds{\alpha_c = 0.1}$\textbf{)}}
& \multicolumn{1}{c}{\textbf{prob (case 2:} $\bolds{\alpha_c = 5}$\textbf{)}}
& \multicolumn{1}{c@{}}{\textbf{prob (case 3:} $\bolds{\alpha_c = 10}$\textbf{)}} \\
\hline
1 & 1 & 1 & 0.989 & 0.989 & 0.981 \\
2 & 2 & 2 & 0.954 & 0.945 & 0.935 \\
3 & 3 & 3 & 0.933 & 0.968 & 0.955 \\
4 & 4 & 4 & 0.950 & 0.980 & 0.967 \\
5 & 5 & 5 & 0.783 & 0.924 & 0.947 \\
\hline
\end{tabular*}
\end{table}

\begin{table}
\caption{Posterior summaries of $c_0$ and $c_1$ for 2VLW ($X$) and
1FAS ($Y$) for three different prior settings} \label{tabbetasse5}
\begin{tabular*}{\tablewidth}{@{\extracolsep{\fill}}lccc@{}}
\hline
\textbf{Parameter} & \textbf{Case 1} & \textbf{Case 2} & \textbf{Case 3} \\\hline
$c_0$ & 1.14 $(0.76,1.74)$ & 1.06 $(0.75,1.56)$ & 1.06 $(0.75,1.47)$ \\
$c_1$ & 1.71 $(1.45,3.30)$ & 1.64 $(1.38,2.14)$ & 1.61 $(1.34,1.91)$ \\
\hline
\end{tabular*}
\end{table}

\begin{table}
\tablewidth=305pt
\caption{Matches from an alignment of the secondary structures of 2VLW
($X$) and 1M9Z ($Z$) for three different prior settings}\label{tabbetasse6}
\begin{tabular*}{\tablewidth}{@{\extracolsep{\fill}}lccccc@{}}
\hline
\textbf{Match} & $\bolds{x}$ & $\bolds{y}$ & \textbf{prob (case 1)}
& \textbf{prob (case 2)} & \textbf{prob (case 3)} \\
\hline
1 & 1 & 1 & 0.958 & 0.990 & 0.981 \\
2 & 2 & 4 & 0.871 & 0.988 & 0.907 \\
3 & 3 & 5 & 0.919 & 0.997 & 0.920 \\
4 & 4 & 6 & 0.936 & 0.990 & 0.957 \\
5 & 5 & 9 & 0.882 & 0.957 & 0.923 \\
\hline
\end{tabular*}
\end{table}

\begin{table}
\tablewidth=305pt
\caption{Posterior summaries of $c_0$ and $c_1$ for 2VLW ($X$) and
1M9Z ($Z$) for three different prior settings} \label{tabbetasse7}
\begin{tabular*}{\tablewidth}{@{\extracolsep{\fill}}lccc@{}}
\hline
\textbf{Parameter} & \textbf{Case 1} & \textbf{Case 2} & \textbf{Case 3} \\
\hline
$c_0$ & 0.82 $(0.68,1.21)$ & 0.82 $(0.70,1.15)$ & 0.83 $(0.70,1.15)$ \\
$c_1$ & 1.18 $(1.08,3.13)$ & 1.17 $(1.09,1.41)$ & 1.17 $(1.08,1.37)$ \\
\hline
\end{tabular*}
\end{table}

For the second protein example, the pair 2VLW-1M9Z, the matches
obtained are shown in Table \ref{tabbetasse6} and the effect on the
parameters $c_0$ and $c_1$ can be seen in Table \ref{tabbetasse7}.
For both pairs, the matching probabilities are generally lower in case
1 (when the prior information on $c$ is weak), although the overall
alignment is still good. Results are robust for larger values of
$\alpha_c$, and further results (not shown) show that the results
remain robust for even larger values of $\alpha_c = \lambda_c$, with
the posterior values of $c$ moving slightly closer to the prior mean of
$1$; this is to be expected as the prior variance gets smaller,
resulting in a more informative prior. However, the results change by
only a small amount, suggesting that the data carry a lot of information.

The implementation of our method does not come with a particularly high
computational cost. The most computationally expensive aspect of our
examples, the unlabeled two-scale model, was implemented in C$++$ and ran
in $10$ seconds on a desktop PC with a 3.10 GHz processor.

\section{Discussion}\label{secdiscussion}

In this paper we have presented a Bayesian model for the pairwise
alignment of two point configurations under full similarity
transformation. The fully Bayesian approach allows for uncertainty in
the transformation parameters to be correctly propagated, which is a
key conceptual difference between our method and others. We note that
isotropic errors have been assumed throughout, but this has been
standard practice in shape analysis [\citet{Shape}]; \citet{TW} have
considered nonisotropic errors in the case of labeled landmarks. Our
emphasis here has been on both the labeled and unlabeled cases.

The work presented here has concentrated on the pairwise alignment of
two configurations. \citet{ruffieux} generalized the method of \citet
{GM} to develop a fully Bayesian model for the alignment of multiple
configurations under rigid body transformations; a natural extension
might therefore be to incorporate our methodology developed in this
paper within their model. \citet{mardia10} addressed the problem of
multiple alignment under rigid body transformations by embedding a
pairwise alignment method within a multi-stage algorithm, and their
methodology could easily be adapted to incorporate the extension to the
full similarity shape case introduced here.

Finally, an important part of our work is the novel application to the
alignment of proteins, using a representation based on secondary
structure elements. This application required the development of our
model to handle more than one scaling factor, since different elements
may require different scaling. The use of one global scaling factor has
been standard practice in shape analysis [\citet{Shape}]. We find that
two scaling factors is sufficient for our needs; our proteins have only
a small number of points, and including more scaling parameters would
come at the cost of overfitting, which our results suggest is
unnecessary. However, our method could be readily extended to include
more scaling factors as needed. This would introduce issues such as
model choice and comparison, and such matters are left for future work.

\begin{appendix}\label{app}
\section*{Appendix: Exponential family and normal approximations}

\subsection{A normal approximation}

Here we give a normal approximation for the exponential family of
distributions, motivated by our requirement for an efficient
proposal\vadjust{\goodbreak}
distribution for the Metropolis method described in Section \ref
{ssecsccond}. Consider the curved exponential family for a continuous
random variable $X$ with density
\[
f(x; \theta) = \exp\bigl\{ a_1 (\theta) b_1 (x) +
a_0(\theta) + b_0(x) \bigr\}.
\]
The second log derivative with respect to $x$ is
\[
\ell''(x) = \partial^2 \log f(x, \theta)/
\partial x^2 = a_1(\theta) b''_1(x)
+ b''_0(x).
\]
We assume that the family is convex so that there is a single mode at
$x = \hat{x}$ and $-\ell'' (\hat{x}) > 0$ uniformly. Then for large
$a_1(\theta)$, we postulate that
%
\begin{equation}
\label{eqnormapprox1}
X \simeq N \bigl(\hat{x}, - \bigl\{ a_1 (
\theta) b''_1 (\hat{x}) +
b''_0 (\hat{x}) \bigr\}^{-1}
\bigr),
\end{equation}
where $\hat{x}$ is the mode of the distribution. A heuristic
explanation follows intuitively using the exchangeability of $x$ and
$\theta$.
For the maximum likelihood estimate $\hat{\theta}$ of~$\theta$, it
is well known that for a large sample size $n$ we have
$ \hat{\theta} \simeq N(\theta,I(\theta)^{-1}), $
where $ I(\theta) $ is the Fisher information, $\mathbb{E}_{\theta
} [-\partial^2 l(\theta;x) /\partial\theta^2 ]$, and
$l(\theta;x)$ is the log-likelihood function.
Consider now interchanging the roles of $x$ and $\theta$, treating
$\theta$ as a variable and $x$ as a parameter. Since $x$ and $\theta$
are exchangeable by conjugacy, we may write
\[
X \simeq N \bigl(\hat{x},- \bigl\{\partial^2 \log f(x, \theta)/
\partial x^2 \bigr\}^{-1}_{x=\hat{x}} \bigr),
\]
which is equivalent to (\ref{eqnormapprox1}) and hence gives a
heuristic demonstration of the result. The validity of this
approximation is confirmed below in various cases where a normal
approximation is well known. [Note that in the case of the normal
distribution with mean $\mu$ and variance $\sigma^2$ the
approximation is exact as required, with $X \sim N(\mu,\sigma^2)$.]

\subsubsection{Gamma}
Consider the gamma distribution with density $p(x) = \beta^{\alpha
}\*x^{\alpha-1}\exp(-\beta x)/\Gamma(\alpha)$. We have\vspace*{2pt} $\ell=
(\alpha-1)\log x -\beta x + \alpha\log\beta-\log\Gamma(\alpha) $
and the mode is $\hat{x} = \frac{\alpha-1}{\beta} $, giving the
approximation
$ X \simeq N (\frac{\alpha-1}{\beta},\frac{\alpha-1}{\beta
^2} ). $
The standard approximation is $X \simeq N (\frac{\alpha}{\beta
},\frac{\alpha}{\beta^2} )$, so the two are approximately the
same for large~$\alpha$.

\subsubsection{Von Mises}
For the von Mises distribution, we have $f(x, \mu) = K\* \exp\{ \kappa
\cos(x-\mu) \}, 0 < x, \mu< 2 \pi$. The mode is $\hat{x} =
\mu$; thus, $(\ell'')_{\hat{x} = \mu} = -\kappa$, and the
approximation is $X \simeq N ( \mu, \frac{1}{\kappa} ),
$ which is a well-known normal approximation to the von Mises
distribution [\citet{mardiajupp}, page 38].

\subsubsection{Halfnormal-gamma}\label{appsechng} For the halfnormal-gamma
distribution, we have
\begin{eqnarray*}
\ell&=& \log f(x;r,\nu,\delta) \\
&=& \log K + (r-1)\log x -\frac{1}{2}\nu
x^2 + \delta x,
\\
\ell' &=& \frac
{(r-1)}{x} - \nu x + \delta
\end{eqnarray*}
and
\[
\ell'' = -\frac
{(r-1)}{x^2} - \nu,
\]
leading to an approximate variance given by $ \{ \nu+ \frac
{(n-1)}{\hat{x}^2} \}^{-1}$. Recall that the mode is $\hat{x} = \{
\delta+ \sqrt{\delta^2 + 4(r-1)\nu} \} / 2\nu$. We therefore have the
approximation $X \simeq N(\hat{x},\operatorname{Var}(X))$. We find the
approximation to be better for larger $r$ and $\delta$; even for small
$r$, the approximation is good for positive values of $\delta$, but
less good for relatively large negative values of $\delta$. Further
details are given in the supplementary material.
\end{appendix}

\section*{Acknowledgments}

Fallaize acknowledges EPSRC funding for his research studies. We thank
Peter Green for helpful comments, and the Editor, Associate Editor and
anonymous referee for their comments which helped to improve a previous
version of the paper.

\begin{supplement}
\stitle{Simulation methods and a normal approximation for the
halfnormal-gamma distribution}
\slink[doi]{10.1214/12-AOAS615SUPP} 
\sdatatype{.pdf}
\sfilename{aoas615\_supp.pdf}
\sdescription{We describe an acceptance-rejection method for simulating
from the halfnormal-gamma distribution and investigate its efficiency
over a range of parameter settings. We also investigate further the
normal approximation to the halfnormal-gamma distribution, which we use
to obtain efficient proposals in our Metropolis updates. We show that
the approximation is best for parameter values where the
acceptance-rejection method is less efficient, and hence that the two
simulation methods complement each other well.}
\end{supplement}


\printaddresses

\end{document}